\documentclass[%
 reprint,
 amsmath,amssymb,
 aps,
 pra
]{revtex4-2}

\usepackage{graphicx}
\usepackage{dcolumn}
\usepackage{bm}
\usepackage{mathtools}
\mathtoolsset{showonlyrefs,showmanualtags}
\usepackage{color}

\newcommand{\reserve}[1]{}

\newcommand{\argmax}{\mathop{\rm arg~max}\limits}

\newcommand{\dif}{\mathrm{d}}
\newcommand{\ecoli}{{\it E. coli} }
\newcommand{\opt}{\mathrm{OPT}}
\newcommand{\Fopt}{F_{\opt}}

\newcommand{\Fexp}{F_{\mathrm{EXP}}}

\newcommand{\e}{\mathrm{e}}
\newcommand{\Prob}{\mathbb{P}}
\newcommand{\Expect}{\mathbb{E}}
\newcommand{\KL}{\mathcal{D}_{\mathrm{KL}}}
\newcommand{\Transpose}{\mathbb{T}}

\newcommand{\Rc}{\mathcal{R}}
\newcommand{\control}{\mathrm{c}}
\newcommand{\uncontrol}{0}
\newcommand{\filter}{\mathrm{f}}
\newcommand{\ccost}{\mathcal{C}}

\newcommand{\nddis}{\tilde{\mathcal{I}}}
\newcommand{\util}{\mathcal{J}}
\newcommand{\ndutil}{\tilde{\mathcal{J}}}

\newcommand{\abar}{\bar{a}}

\newcommand{\eqnref}[1]{Eq. \eqref{#1}}

\newcommand{\pos}{\xi}
\newcommand{\motor}{M}
\newcommand{\dir}{X}
\newcommand{\obs}{Y}
\newcommand{\post}{Z}

\newcommand{\W}{W}
\newcommand{\gain}{K}

\newcommand{\ndC}{\tilde{\lambda}}
\newcommand{\ndb}{\tilde{\beta}}
\newcommand{\ndV}{\tilde{V}}
\newcommand{\ndR}{\tilde{R}^{\uncontrol}}

\newcommand{\ndt}{\tau}

\newcommand{\const}{I}

\begin{document}

\preprint{APS/123-QED}

\title{Optimal sensing and control of run-and-tumble chemotaxis}

\author{Kento Nakamura$^1$}
\author{Tetsuya J. Kobayashi$^1$}%
\altaffiliation[Also at ]{Institution of Industrial Science, the University of Tokyo, 4-6-1, Komaba, Meguro-ku, Tokyo, 153-8505, Japan}
\altaffiliation[Also at ]{Universal Biology Institute, The University of Tokyo, 7-3-1, Hongo, Bunkyo-ku, 113-8654, Japan.}
 \email{tetsuya@mail.crmind.net}
\affiliation{%
$^1$Department of Mathematical Informatics, the Graduate School of Information Science and Technology, the University of Tokyo, 7-3-1, Hongo, Bunkyo-ku, Tokyo, 113-8654, Japan,
}%



\date{\today}

\begin{abstract}
Run-and-tumble chemotaxis is one of the representative search strategies of an odor source via sensing its spatial gradient.
The optimal ways of sensing and control in the run-and-tumble chemotaxis have been analyzed theoretically to elucidate the efficiency of strategies implemented in organisms.
However, because of theoretical difficulties, most of attempts have been limited only to either linear or deterministic analysis even though real biological chemotactic systems involve considerable stochasticity and nonlinearity in their sensory processes and controlled responses. 
In this paper, by combining the theories of optimal filtering and Kullback-Leibler control of partially observed Markov decision process (POMDP), we derive the optimal and fully nonlinear strategy for controlling run-and-tumble motion depending on noisy sensing of ligand gradient.
The derived optimal strategy consists of the optimal filtering dynamics to estimate the run-direction from noisy sensory input and the control function to regulate the motor output.
We further show that this optimal strategy can be associated naturally with a standard biochemical model and experimental data of the \textit{Escherichia coli}'s chemotaxis.
These results demonstrate that our theoretical framework can work as a basis for analyzing the efficiency and optimality of run-and-tumble chemotaxis.
\end{abstract}

\maketitle


\section{Introduction}
A wide variety of organisms, from animals to single cells, exhibit abilities to search for odor sources.
The abilities are essential for obtaining food, suitable environments, and mates, and it is expected that the searching strategies of organisms are subject to some selection pressures and optimized evolutionarily.
Thus, it has been explored how efficiently or optimally searching strategies of various living organisms are designed \cite{benichou2011intermittent,hein2016natural,baker2018algorithms}.
One of the strategies investigated the most intensively is a run-and-tumble motion in chemotaxis of \textit{Escherichia coli}, whose biochemical signaling pathways have also been elucidated \cite{berg2004coli}.

\ecoli climbs spatial gradients of ligand concentration by sensing temporal changes of ligand concentration and regulating the motor accordingly.
The motor switches between run and tumble states, resulting in repeated ballistic swimming (run) interrupted with random changes of direction (tumble).
By inhibiting tumble when sensing an increase of ligand concentration and vice versa, \ecoli can selectively enhance the positive displacement along the increasing direction of the gradient.
Such sensory-motor cycle in \ecoli's chemotaxis has been theoretically modeled \cite{vladimirov2008dependence,jiang2010quantitative,celani2011molecular,si2012pathway,dufour2014limits} based on the biochemical model of the intracellular signaling pathway \cite{bray1998receptor,barkai1997robustness,yi2000robust,sourjik2004functional,keymer2006chemosensing,mello2007effects,tu2008modeling}.
\begin{figure*}
    \centering
    \includegraphics[width=\linewidth]{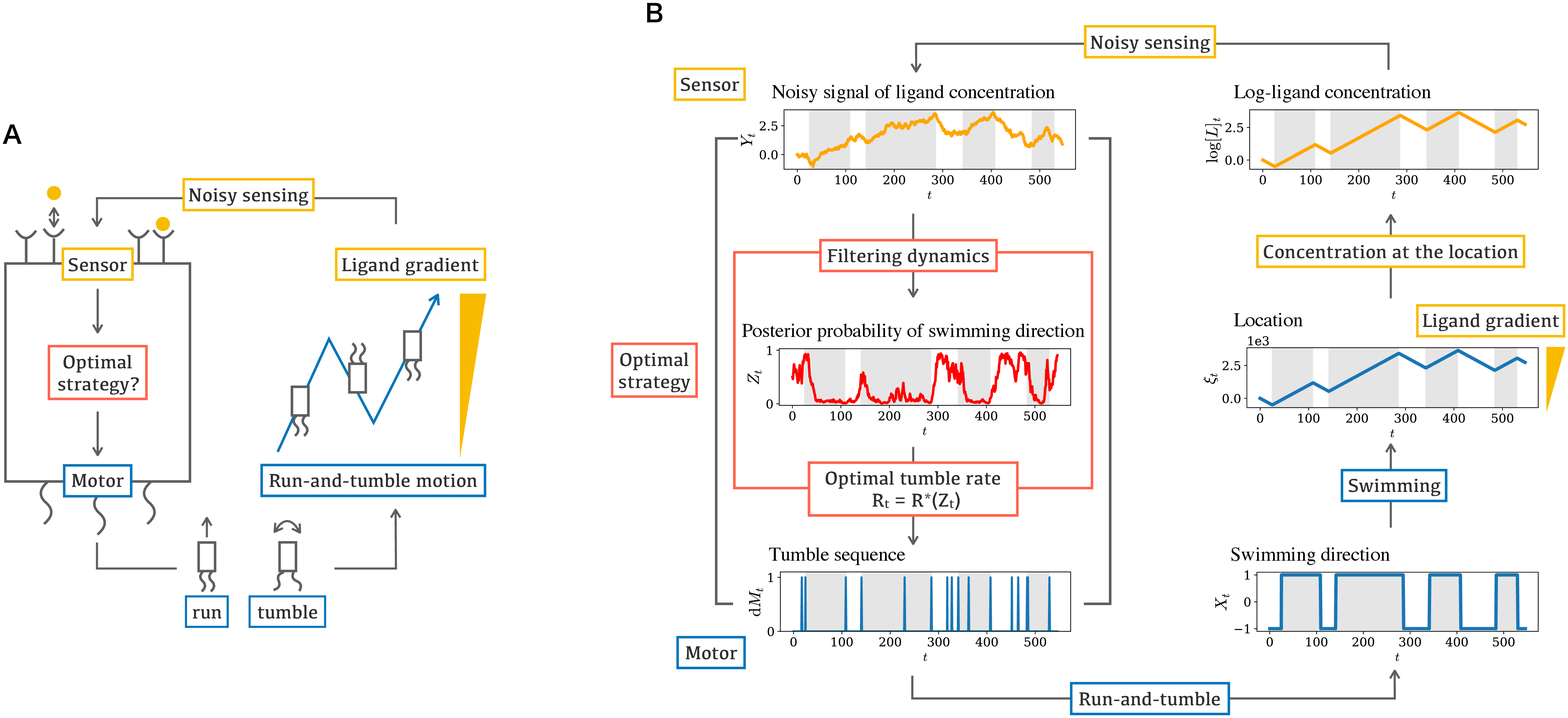}
    \caption{(A)Schematic diagram of the sensory-motor cycle in \ecoli's chemotaxis. (B) Sample paths of variables involved in the cycle driven by the optimal strategy.
    Each panel shows tumble events sequence, $\motor_{t}$, swimming direction, $\dir_{t}$, location, $\xi_{t}$, log-ligand concentration, $\log[L]_{t}$, sensing signal, $\obs_{t}$, and the posterior probability of the swimming direction, $\post_{t}$.
    Gray regions represent the time intervals during which the swimming direction is up the gradient, i.e., $\dir_{t}=+1$.
    Those trajectory are simulated by using the dimensionless parameter values, $\ndb=10^{0.6}$, $\ndR=10^{0.4}$, and $\ndC=10^{0.5}$. 
    In the context of \ecoli chemotaxis, these dimensionless parameter values can correspond to the following biologically-realistic dimensional parameter values: $v = 20\mathrm{\mu m\cdot s^{-1}}$, $c=10^{-3} \mathrm{\mu m^{-1}}$, $\sigma=
    8.7\times 10^{-3}\mathrm{s^{-1}}$, $\gamma=9.2\times 10^{-3} \mathrm{s^{-1}}$, $\beta=1.8\times 10^{-3}\mathrm{\mu m^{-1}}$, and $R^{\uncontrol}=2.3\times 10^{-2} \mathrm{s^{-1}}$.
    }
    \label{fig:sensory-motor cycle}
\end{figure*}

To discuss how well \ecoli is structured to perform chemotaxis, it is important to understand its sensory-motor cycle in terms of the optimality.
While biochemical models describe how the actual biochemical pathway is organized, optimality models aim to characterize the best possible chemotactic performance under physical constraints and the strategy that achieves the performance.
By comparing the optimality models with \ecoli's chemotactic response measured experimentally \cite{block1982impulse,block1983adaptation,segall1986temporal,sourjik2002receptor,sourjik2004functional,shimizu2010modular}, we can discuss how closely to the physical limit a cell can perform chemotaxis and whether its biochemical signaling pathway is organized in a reasonable way for chemotaxis.
Previous attempts using optimality models can be classified into two directions.
One uses optimal control models, which describe the optimal strategy to modulate the tumble rate based on sensing signals.
This approach formulates the tumble rate as a functional of sensing signals and considers the optimization of the functional with respect to chemotactic performance such as gradient climbing \cite{de2004chemotaxis,clark2005bacterial,celani2010bacterial,Mattingly2021.02.22.432091}.
The other uses optimal filtering models.
Since sensory noise is not ignorable at the cellular level, \ecoli's biochemical signaling pathway is expected to possess noise immune properties.
By deriving the optimal filtering strategy that extracts ligand concentration or its changes from the noisy signal, we can discuss how well \ecoli's biochemical signaling pathway is designed to decode information \cite{andrews2006optimal,mora2010limits,aquino2014memory,mora2019physical,nakamura2021connection}.

Despite the progress in optimality models, there is still no theory that can properly deal with "sensory-motor cycle", "nonlinear response" and "sensory noise" at the same time from the perspective of optimality.
The optimality of the whole sensory-motor cycle is crucial because the sensing strategy shapes the information available for motor control and the motor control strategy affects what would be sensed through the modulation of a cell motion.
However, previous optimal filtering models \cite{andrews2006optimal,mora2010limits,aquino2014memory,mora2019physical,nakamura2021connection} neglect the influence of motor control on sensing signals.
In addition to motor control, previous optimality models fail to include the nonlinear responses observed for example in the biochemical pathway of \ecoli chemotaxis \cite{tu2008modeling,shimizu2010modular}.
They consider only linear responses and the optimal strategy in that class by assuming a weak ligand gradient\cite{de2004chemotaxis,clark2005bacterial,celani2010bacterial,Mattingly2021.02.22.432091}.
Also, sensory noise is neglected in the derivation of those optimal control models.
There are several theoretical studies that partly deal with the optimality of sensory-motor cycle under the existence of nonlinearity and/or noise \cite{kollmann2005design,kafri2008steady,celani2011molecular,flores2012signaling,he2016noise,hein2016physical,long2017feedback,dev2018optimal,kirkegaard2018role,gosztolai2020cellular,mandal2021effect}.
However, it is generally difficult to analytically investigate the effects of noise and nonlinearity, and therefore such effects are often treated only by simulations.
A few of those studies analytically treat the impact of noise and nonlinearity on the optimality but do it at the parameter level by assuming fixed structures of strategies.
To discuss whether the actual biochemical pathway of \ecoli is structurally optimized or not, it is important to include a sufficiently large class of strategies as candidates of an optimal strategy.

In this article, we propose an optimality model that can take into account "sensory-motor cycle", "nonlinear response", and "sensory noise" integratively based on the framework of partially observable optimal control \cite{segall1977optimal}.
 We derive the optimal strategy for tumble rate regulation and show that the strategy is implemented as a combination of two separated components: an optimal filtering dynamics that yields an estimate of the swimming direction and an optimal control function that converts the estimate to tumble rate.
We show that the obtained optimal strategy can be related to both a standard biochemical model and experimental data of \ecoli's signaling pathway.
Finally, we discuss possible extensions of the optimality model.

\section{Modeling and Formulation}
To consider a minimal setting, we focus on the control strategy to regulate run-and-tumble motion on one-dimensional axis along a spatial ligand gradient.
We model a tumble as an instantaneous event which occurs according to a Poisson process with rate $R_{t} \geq 0$ and denote by $M_{t}\in\mathbb{N}$ the total number of tumble events up to time $t$.
We ignore the dispersal of a swimming direction during the run phase and assume that \ecoli changes its direction only at tumble events.
We denote by $\dir_{t}\in\{+1,-1\}$ the swimming direction and assume that $\dir_{t}$ flips with probability $1/2$ at every time when tumble occurs (See SM for a more general case where the directional change occurs with other probabilities).
With these assumptions, we can express the time evolution of the probability of $X_{t}$ under a given tumble rate $R_{t}$ by a continuous-time Markov chain:
\begin{align}
    \frac{\dif \bm{p}_{t}}{\dif t} = \frac{1}{2} R_{t}\left(\begin{array}{cc}
        -1 & 1 \\
        1 & -1
    \end{array}\right)\bm{p}_{t} \label{eq:state},
\end{align}
where $\bm{p}_{t} := (\Prob(\dir_{t}=+1),\Prob(\dir_{t}=-1))^{\Transpose}$ and the initial condition is set to $\Prob(\dir_{t}=-1)=\pi\in[0,1]$.
We denote by $\pos_{t}\in\mathbb{R}$ the position of \ecoli and assume that \ecoli moves at a constant speed $v$ to the current direction, $\dif \xi_{t}/\dif t = v\dir_{t}$.
Note that if the tumble rate, $R_{t}$, is constant, there is no net displacement on average because the transition matrix of the direction is symmetric.

Under this setting, we consider the optimal strategy for regulating the tumble rate, $R_{t}$, to climb a spatial ligand gradient depending on the noisy sensing of environment.
We introduce sensory noise by defining a sensing signal, $\obs_{t}$, and specify the available information for the regulation of $R_{t}$ as follows:
\begin{align}
    R_{t} =& \Rc^{\control}_{t}[\obs_{0:t}], \label{eq:control_candidate}\\
    \obs_{t} :=& h(\pos_{t}) + \sqrt{\sigma}W_{t}, \label{eq:sensing_signal}
\end{align}
where $\obs_{0:t}:=\{\obs_{t'}\mid t'\in[0,t]\}$. $\Rc^{\control}_{t}$ is a control functional representing the strategy of tumble regulation, which is allowed to depend on time explicitly.
$h$ is a function that we specify later.
$W_{t}$ is a standard Wiener process, and $\sigma>0$ represents the noise intensity.
Note that we do not restrict the candidates of strategies only to those being realized by a linear response.
Of all possible control functionals, we aim to find the optimal control functional, $\Rc^{\ast}_{t}$, that satisfies the following maximization problem:
\begin{align}
    \Rc^{\ast}_{0:\infty}:=\argmax_{\Rc^{\control}_{0:\infty}} \util(\pi,\Rc^{\control}_{0:\infty}), \label{eq:maximization_problem}
\end{align}
where $\util$ is a utility functional defined as follows:
\begin{align}
    &\util(\pi,\Rc_{0:\infty}^{\control}) := \Expect^{\control}\left[\int_{0}^{\infty}\e^{-\gamma t}v\dir_{t}\dif t \right] - \frac{1}{\beta}\ccost[\Rc^{\control}_{0:\infty}].\label{eq:utility_function}
\end{align}
Here, we introduce a control cost, $\ccost$, to obtain a bounded control functional and take into account a possible physical cost of tumble regulation. 
$\Expect^{\control}$ indicates that the expectation is taken under controlled process generated by \eqnref{eq:state}, \eqref{eq:control_candidate}, and \eqref{eq:sensing_signal}.
$1/\beta>0$ and  $\gamma>0$ represent the weight of control cost and a temporal discounting rate, respectively.

Defining the control cost is nontrivial, especially for biological systems. While quadratic cost functions are conventionally used in linear control, a quadratic form of  $\ccost[\Rc^{\control}_{0:t}]$ may not admit a natural interpretation.
In addition, $\ccost[\Rc^{\control}_{0:t}]$ should be designed appropriately to guarantee that the derived optimal control functional does not produce biologically irrelevant values such as $R_{t} <0$ or $R_{t}=0$.
To deal with this difficulty, we introduce a control cost by using the framework of Kullback-Leibler (KL) control.
KL control framework assumes an uncontrolled process
and define the control cost by the discrepancy of the controlled process from the uncontrolled one in terms of the KL divergence, $\KL[\Prob^{\control}\parallel\Prob^{\uncontrol}]$, where $\Prob^{\control}$ and $\Prob^{\uncontrol}$ denote the path probabilities of the controlled and uncontrolled processes, respectively.
We model an uncontrolled process of tumbling by a reference constant-rate process, i.e., $R_{t}^{\uncontrol} = R^{\uncontrol} > 0$.
Because the constant rate Poisson process is the simplest and the least predictable, it can work as the representative of the uncontrolled situation. 
With a temporal discounting factor, we derive the control cost functional under our setting as (see SM for derivation):
\begin{align}
    \ccost[\Rc_{0:\infty}^{\control}]:=\Expect^{\control}\left[\int_{0}^{\infty}\e^{-\gamma t}\left\{\log\frac{R_{t}}{R^{\uncontrol}}\dif M_{t} - (R_{t}-R^{\uncontrol})\dif t\right\}\right].
    \label{eq:KL_control_cost}
\end{align}

We can derive the optimal control functional, $\Rc^{\ast}_{0:\infty}$ in \eqnref{eq:maximization_problem} based on the framework of partially observable control \cite{segall1977optimal} (See Supplementary Material (SM) for derivation).
Even though we assume that the tumble rate $R_{t}$ at each time $t$ can depend on the entire history of the sensing signal, $\obs_{0:t}$, 
we can prove that the optimal control under a certain condition is achieved only by using the posterior probability of the direction $\post_{t} := \Prob(\dir_{t}=-1\mid \obs_{0:t})$, which summarizes the information contained in $Y_{0:t}$.
In particular, if the function $h$ in sensing signal (\eqnref{eq:sensing_signal}) is affine, $h(\xi_{t}) = \lambda\xi_{t}+const.$, where $\lambda$ is a constant, the optimal control functional is expressed by an optimal control function, $R^{\ast}$, which takes $\post_{t}$ as its argument:
\begin{align}
    \Rc^{\ast}_{t}[\obs_{0:t}] =& R^{\ast}(\post_{t}) := R^{\uncontrol}\exp\left\{-\ndb\left(\post_{t}-\frac{1}{2}\right)\frac{\dif \tilde{V}}{\dif \post}(\post_{t}) \right\}.\label{eq:optimal_control_law}
\end{align}
Here, $\tilde{V}$ is a scaled value function on $[0,1]$ which satisfies $\tilde{V}(\pi) = \max_{\Rc_{0:\infty}^{\control}}\ndutil(\pi,\Rc_{0:\infty}^{\control})$, where $\ndutil := (\gamma/v) \util$ is a scaled utility functional.
$\tilde{V}$ is obtained by solving the following second-order partial differential equation known as the Hamilton-Jacobi-Bellman equation in the optimal control theory:
\begin{align}
    \tilde{V} =& -\frac{\ndR}{\ndb} \left[\exp\left\{-\ndb\left(\post-\frac{1}{2}\right)\frac{\dif \tilde{V}}{\dif \post} \right\}-1\right]\\
    &\quad + \{\ndC\post(1-\post)\}^2\frac{\dif^2 \tilde{V}}{\dif \post^2} + (2\post-1),\label{eq:optimal_bellman_equation}
\end{align}
where $\ndR:=R^{\uncontrol}/\gamma$, 
$\ndC := \sqrt{2}\lambda v/\sqrt{\sigma\gamma}$, and $\ndb:= \beta v/\gamma$ are dimensionless parameters, which describe the ratio between the temporal-discounting rate and the reference tumble rate, the signal-to-noise ratio of the sensory signal $\obs_{t}$, and a relative weight of the net displacement with respect to the control cost, respectively.


We can show that the information needed for the optimal control, $\post_{t}$, can be calculated sequentially and separately before feeding into the optimal control function (\eqnref{eq:optimal_control_law}).
Based on the theory of nonlinear filtering, we derive the time evolution of $\post_{t}$ as a following stochastic differential equation:
\begin{align}
    \frac{\mathrm{d}\post_{t}}{\mathrm{d}t}=&- R^{\filter}(\post_{t})\left(\post_{t}-\frac{1}{2}\right)-\gain\post_{t}(1-\post_{t})\circ\frac{\mathrm{d}\obs_{t}}{\mathrm{d}t}, \label{eq:filter_expectation}
\end{align}
which holds when a initial value is set, $\post_{0}=\pi$, a function $R^{\filter}(\cdot)$ is set to $R^{\ast}(\cdot)$, and a parameter $K$ is adjusted as $K = 2\lambda v/\sigma$.
Here, $\circ$ represents the Stratonovich integral.
The first and the second terms represent the prediction based on a prior knowledge about directional change represented by \eqnref{eq:state} and the update of the posterior based on the sensing signal (\eqnref{eq:sensing_signal}), respectively \cite{nakamura2021connection}.

From the above discussion, we obtain the optimal strategy for tumble regulation as the combination of the filtering dynamics (\eqnref{eq:filter_expectation}) of the sensing signal and the optimal control function (\eqnref{eq:optimal_control_law}) driven by the filtered signal $Z_{t}$.



\section{Comparison and correspondence with biochemical model}
To relate the optimal but abstract strategy represented by \eqnref{eq:filter_expectation} and \eqnref{eq:optimal_control_law} to the actual \ecoli’s biochemical pathway, we introduce additional two assumptions about the sensing process.
First, we assume that the ligand concentration, $[L]_{t}$, sensed by \ecoli is exponentially distributed with the position, $[L]_{t}\propto \exp(c\xi_{t})$, where $c$ represents the steepness of the gradient.
Second, we assume that the sensory noise is Gaussian in log-concentration, $\obs_{t} = \log[L]_{t}+\sqrt{\sigma}\W_{t}$.
The setting specified by these two assumptions satisfies the condition that $h$ in the sensing signal (\eqnref{eq:sensing_signal}) is affine with the parameter set to $\lambda=c$.
Thus, the optimal strategy is given by \eqnref{eq:filter_expectation} and \eqnref{eq:optimal_control_law}.


Under this setting, we transform \eqnref{eq:filter_expectation} to the form which clarifies the relation with the standard biochemical model of \ecoli's signaling pathway \cite{tu2008modeling}.
By following the previous study \cite{nakamura2021connection}, we define a log-likelihood ratio by $\theta_{t} := \log \{(1-\post_{t})/\post_{t}\}$ and introduce a variable, $\mu_{t}$, which we call a prediction term.
Then, we can derive an equivalent expression of the filtering dynamics \eqnref{eq:filter_expectation}:
\begin{align}
    \post_{t} =& \frac{1}{1+\exp(\theta_{t})}, \label{eq:fd_posterior}\\
    \theta_{t} =& - \kappa \mu_{t}+\gain\left[ \log [L]_{t} + \sqrt\sigma \W_{t}\right]+\phi,\label{eq:fd_likelihood-ratio}\\
    \frac{\dif \mu_{t}}{\dif t} =& -\frac{R^{\filter}(\post_{t})}{\kappa} \frac{\post_{t}-1/2}{\post_{t}(1-\post_{t})}]=:\Fopt(\post_{t}),  \label{eq:fd_prediction}
\end{align}
where $\kappa>0$ is an arbitrary constant and $\phi := \log\{(1-\pi)/\pi\}- K\log[L]_{0}+\kappa\mu_{0}$ is a constant of integral (see SM for derivation).

Next, we briefly review a standard biochemical model of \ecoli’s signaling pathway represented by \eqnref{eq:tu_activity}, \eqref{eq:tu_free-energy}, and \eqref{eq:tu_methylation}, which are subsequently compared with \eqnref{eq:fd_posterior}, \eqref{eq:fd_likelihood-ratio}, and \eqref{eq:fd_prediction}. 
In an \ecoli's cell, the ligand is sensed by an array of receptors, each of which takes either an active or inactive state, and the active receptors increase the rate of tumble via mediating proteins.
The biochemical model defines and quantifies the receptor activity, $a_{t}$, by the ratio of the active receptors.
The receptors are also influenced by their methylation, which is quantified in the model by the average methylation level, $m_{t}$.
The dependence of $a_{t}$ on $m_{t}$ and on the ligand concentration, $[L]_{t}$, is described by the Monod-Wyman-Changeux (MWC) allosteric model:
\begin{align}
    a_{t} =& \frac{1}{1+\exp(f_{t})},    \label{eq:tu_activity}\\
    f_{t} =& N(-\alpha m_{t} + \log [L]_{t} + \const),\label{eq:tu_free-energy}
\end{align}
where $f_{t}$ is the free energy difference between active/inactive states. 
$N,\alpha$, and $\const$ are biochemical parameters.
The receptor activity, $a_{t}$, affects the methylation and demethylation enzymes, CheR and CheB, and the resultant kinetics of methylation is modeled as:
\begin{align}
    \frac{\dif m_{t}}{\dif t} = F(a_{t}), \label{eq:tu_methylation}
\end{align}
where $F$ is a decreasing function on $[0,1]$ with the single zero point, $\abar$.
We can see that the kinetics represented by $F$ works as a negative feedback on $a_{t}$ by noting that $\partial a_{t}/\partial m_{t}>0$ and $F'(a_{t})<0$ hold.
The dependence of tumble rate on the receptors' state can be modeled by setting the tumble rate as an increasing function of $a_{t}$ \cite{cluzel2000ultrasensitive,vladimirov2008dependence,jiang2010quantitative,si2012pathway,dufour2014limits}.
The biochemical model composed of Eqs. \eqref{eq:tu_activity}-\eqref{eq:tu_methylation} can describe the experimental responses of \ecoli to various ligand profiles \cite{tu2008modeling,tu2013quantitative}.

We can now see that the biochemical quantities $(a_{t},f_{t},m_{t})$ in Eqs. \eqref{eq:tu_activity}-\eqref{eq:tu_methylation} have the same structure as the filtering quantities $(\post_{t},\theta_{t},\mu_{t})$ in Eqs. \eqref{eq:fd_posterior}-\eqref{eq:fd_prediction}.
The posterior probability, $\post_{t}$, and the receptor activity, $a_{t}$, are both described by the sigmoidal transformation of $\theta_{t}$ and $f_{t}$, respectively.
The log-likelihood ratio, $\theta_{t}$, and the free energy difference, $f_{t}$, are expressed by the sum of $\log[L]_{t}$, $\mu_{t}$, and $m_{t}$.
The dynamics of the prediction term, $\mu_{t}$, and the methylation level, $m_{t}$, are described by the functions of $\post_{t}$ and $a_{t}$, respectively.
The tumble rate is modeled by the function of $\post_{t}$ and $a_{t}$ in the optimality and biochemical models, respectively.
This correspondence between the optimality and biochemical models suggests that the \ecoli's biochemical signaling pathway possesses the desirable structure for the optimal control of tumble regulation under sensory noise.

\begin{figure}
\centering
    \includegraphics[width=\linewidth]{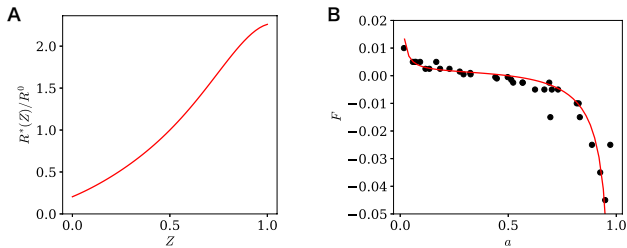}
    \caption{
    (A)A numerical solution of optimal control function $R^{\ast}$.
    (B)The comparison between the feedback function of the optimality model, $F_{\opt}$, (red curve) and that of the experimental data, $F_{\exp}$, (black  points)  \cite{shimizu2010modular}. Parameter values in $R^{\ast}$ and $F_{\opt}$ are set to $R^{\uncontrol}=2.3\times 10^{-2}$, $\ndb=10^{0.6}$, $\ndR=10^{0.4}$, and $\ndC=10^{0.5}$ (see SM for the fitting procedure).}
    \label{fig:F-a}
\end{figure}

To see whether the dependence of tumble rate on $\post_{t}$ and $a_{t}$ is consistent between the optimality and biochemical models, we investigate the functional form of the optimal control function, $R^{\ast}(\post_{t})$.
We calculate $R^{\ast}$ by substituting into \eqnref{eq:optimal_control_law} the value function, $\tilde{V}$, obtained by numerically solving the HJB equation (\eqnref{eq:optimal_bellman_equation}).
In Figure \ref{fig:F-a} (A), we show the optimal control function for a representative set of parameter values.
We can see that $R^{\ast}(\post)$ monotonically increases with $\post$ and is greater than $R^{\uncontrol}$ for $\post>1/2$ and smaller than $R^{\uncontrol}$ for $\post<1/2$.
As we show below in Fig. \ref{fig:optimal_tumble_rate}, this property of $R^{\ast}$ is robust to changes in parameter values (See also SM for the detail of the functional form of $R^{\ast}$).
The result indicates that the tumble rate should be high when the direction is likely to be down the gradient and vice versa to perform chemotaxis optimally under noise.
This property of the optimal control function is consistent with the regulation of tumble rate in \ecoli's cell that can be modeled by an increasing function of $a$ \cite{vladimirov2008dependence,jiang2010quantitative,si2012pathway,dufour2014limits}.

We then clarify whether the optimal feedback function, $\Fopt(\post)=-\{R^{\ast}(\post)/\kappa\}(\post-1/2)/\{\post(1-\post)\}$, in \eqnref{eq:fd_prediction} explains the corresponding feedback function, $F$, in \eqnref{eq:tu_methylation}, which describes the methylation kinetics.
Although the biochemical mechanism governing the kinetics is not fully understood, the functional form of $F$ is experimentally estimated using exponential ramp-response measurements \cite{shimizu2010modular}.
We compare $\Fopt$ with the experimentally measured feedback function, $\Fexp$.
We obtain $\Fopt$ by plugging a numerical solution of $R^{\ast}$ into the equation of $\Fopt$ and fit $\Fopt$ to $\Fexp$ by adjusting $\ndb$, $\ndR$, and $R^{\uncontrol}$ (see SM for detail of fitting procedure).
Fig. \ref{fig:F-a} shows the functional form of $\Fopt$ and the experimental data, $\Fexp$ \cite{shimizu2010modular}.
We can see that $\Fopt$ explains $\Fexp$ well, particularly its characteristic non-linearity; gentle slope around $a=1/2$ and steep slope near $a=1$.
The steep slope near $a=1$ is conjectured to be generated by CheB phosphorylation and its functional role is not fully clarified \cite{kollmann2005design,shimizu2010modular,Keegstra2017-st}.
The agreement between $\Fopt$ and $\Fexp$ indicates that the methylation kinetics represented by $\Fexp$ can be understood as the property of the optimal filtering incorporating the prediction of motor regulation.

We should note that, in our previous work, we related $\Fexp$ to the feedback function obtained without considering the motor control\cite{nakamura2021connection}.
The feedback function there was $\Fopt(z)=- (R/\kappa)(z-1/2)/\{z(1-z)\}$ where $R^{\ast}(z)$ in \eqnref{eq:fd_prediction} is replaced with a constant $R$. 
Because this function is symmetric and has zero at $z=1/2$ whereas $\Fexp$ is asymmetric and has zero near $a\approx 0.3$, a change of variable for $Z$ had to be introduced to fit $\Fopt$ to $\Fexp$ in the previous work.
In this work, the change of variable is no longer necessary because the effect of the control, $R^{\ast}(Z)$, can account for the asymmetry of the function.
Thus, we have obtained a more natural explanation of $\Fexp$ by considering the optimal motor control.

\section{Parameter dependency of control function and performance}
The results described above indicates that the \ecoli's biochemical pathway can be understood as a physical implementation of the optimal strategy under sensory noise.

\begin{figure}
\centering
    \includegraphics[width=\linewidth]{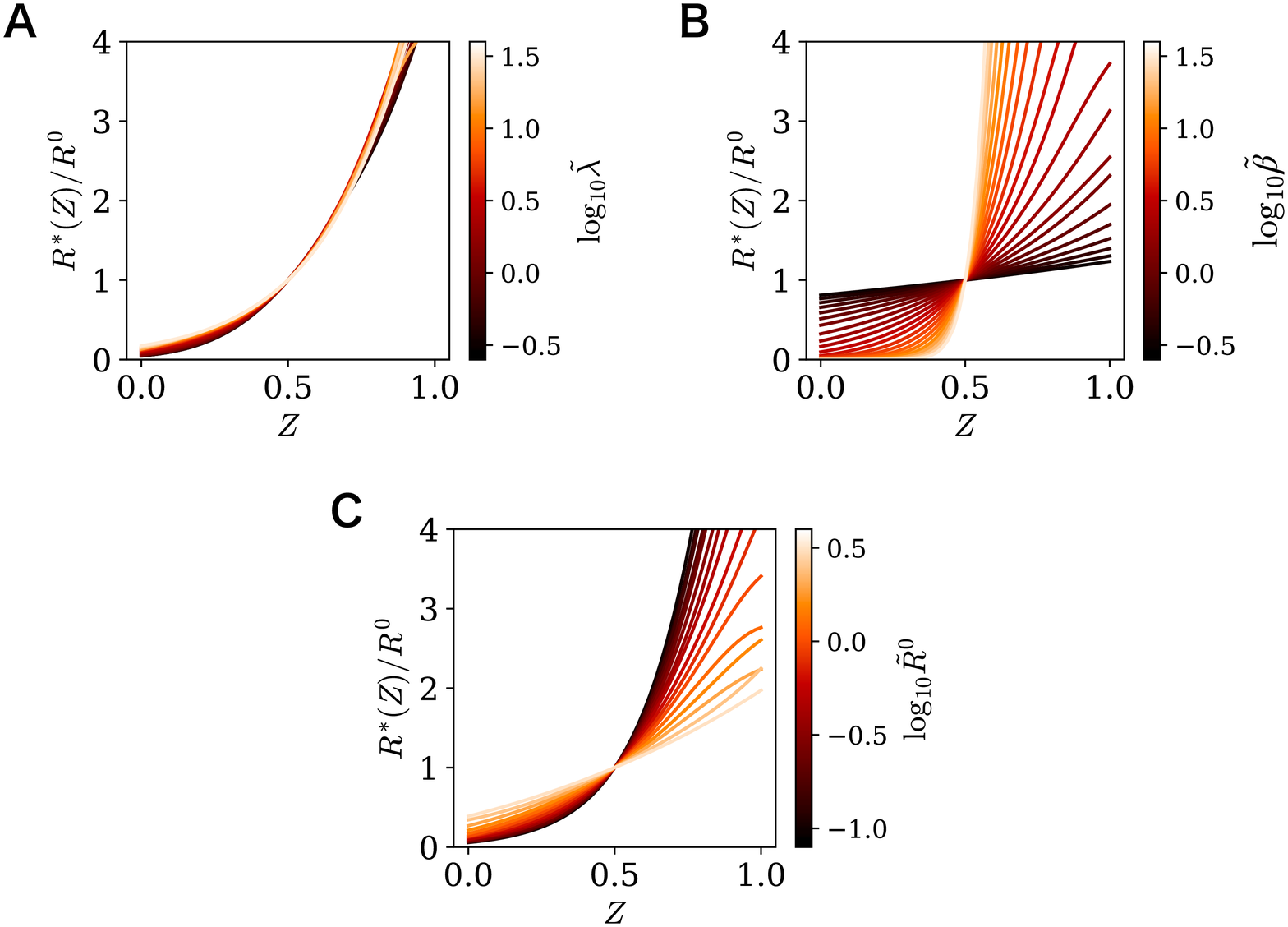}
    \caption{Dependence of normalized optimal control function $R^{\ast}(Z)/R^{\uncontrol}$ on parameters $\ndC$ (A), $\ndb$ (B), and $\ndR$ (C). We used the parameter values, $\ndb=10^{0.5},  \ndR=1,  \ndC=10^{0.5}$, except designated in each panel.}
    \label{fig:optimal_tumble_rate}
\end{figure}

\begin{figure}
    \centering
    \includegraphics[width=\linewidth]{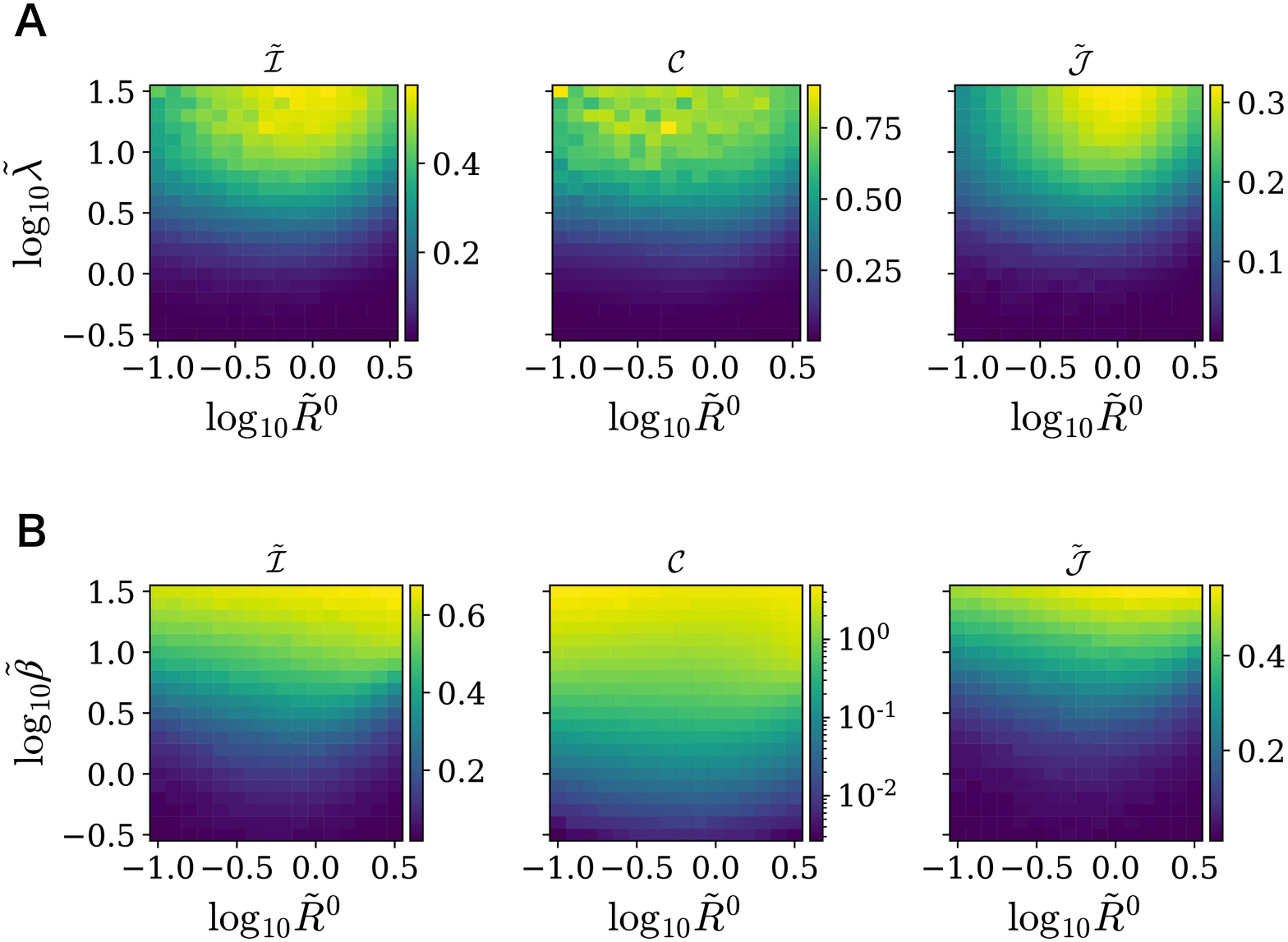}
    \caption{Dependence of the net displacement, $\nddis=\int_{0}^{\infty}\e^{-\ndt}\dir_{\ndt}\dif\ndt$, the control cost, $\ccost$, and the utility, $\ndutil=\nddis-\ccost/\ndb$ on $\ndC$ and $\ndR$ (A) and on  $\ndb$ and $\ndR$ (B). We used the parameter values, $\ndb=10^{0.5},  \ndR=1,  \ndC=10^{0.5}$, except designated in each panel.
    }
    \label{fig:utility}
\end{figure}

We then explore how the optimal control strategy and its chemotactic performance depend on conditions such as the signal-to-noise ratio by numerical simulations under varied parameters.
Of the control function (\eqnref{eq:optimal_control_law}) and the filtering dynamics (\eqref{eq:filter_expectation}) that consist the optimal strategy, we here focus only on the parameter dependence of the normalized optimal control function, $R^{\ast}(\post)/R^{\uncontrol}=\exp\{-\ndb(\post-1/2)\partial \ndV/\partial \post)\}$ (\eqnref{eq:optimal_control_law}) (See ref. \cite{nakamura2021connection} and SM for the parameter dependence of the filtering dynamics (\eqnref{eq:filter_expectation})).
To quantify the chemotactic performance, we introduce three non-dimensional performance indices: the scaled net displacement along the ligand gradient, $\nddis:=\Expect[\int_{0}^{\infty}\e^{-\tau}\dir_{\tau}\dif\tau]$, the control cost, $\ccost$, and the scaled utility, $\ndutil := (\gamma/v)\util=\nddis-\ccost/\ndb$ where we define $\tau:=\gamma t$.
We note that the function, $R^{\ast}(\post)/R^{\uncontrol}$, and the performance indices, $\nddis$, $\ccost$, and $\ndutil$ are determined only by the dimensionless parameters, $\ndb$, $\ndR$, and $\ndC$, which appear in the HJB equation (\eqnref{eq:optimal_bellman_equation}).
We plot the parameter dependence of the normalized optimal control function and the performance indices in Figs. \ref{fig:optimal_tumble_rate} and \ref{fig:utility}, respectively.

We first investigate whether a high signal-to-noise ratio improves the net displacement and how the control strategy changes depending on the non-dimensional signal-to-noise ratio, $\ndC$.
From Fig. \ref{fig:utility}(A), we can see that in a high signal-to-noise ratio situation, a large net displacement can be obtained at the expense of a high control cost, as both the net displacement, $\nddis$, and the control cost, $\ccost$, are larger when $\ndC$ is higher.
On the other hand, Fig. \ref{fig:optimal_tumble_rate}(A) shows that the optimal control function is almost unchanged even though the value of $\ndC$ changes.
We can understand this phenomenon by considering that the signal-to-noise ratio is reflected in tumble regulation through the uncertainty of the estimator.
If the signal-to-noise ratio is high, the swimming direction can be estimated with high precision to be either up $\post_{t}\approx 0$ or down $\post_{t}\approx 1$ the gradient.
If the signal-to-noise ratio is low, in contrast, it is difficult to estimate accurately whether the swimming direction is up or down and thus the estimator tends to take the intermediate value around $\post_{t}\approx 0.5$.
(See also SM for the behavior of $\post_{t}$ under different values of $\ndC$.).
Therefore, even though the functional form of $R^{\ast}/R^{\uncontrol}$ is almost invariant to $\ndC$, the actual tumble rate would be modulated drastically under high $\ndC$ and gently under low $\ndC$, respectively.

We next investigate whether and how the net displacement can be increased by varying the relative weight of the control cost to the net displacement in the utility functional, i.e., $1/\ndb$.
Fig. \ref{fig:utility}(B) shows that both the net displacement, $\nddis$, and the control cost, $\ccost$, are increasing functions of $\ndb$ and Fig. \ref{fig:optimal_tumble_rate}(B) shows that the slope of $R^{\ast}/R^{\uncontrol}$ increases with the increase of $\ndb$.
This result indicates that, as the weight of control cost decreases, the optimal control strategy tends to use abrupt switching of motors to increase the net displacement. 
If the weight of control cost is high, the optimal strategy employs moderate switching of the motor.
Thus, the steepness of the control function may be used to estimate potential control costs behind actual biological system.

We finally investigate, by varying $\ndR=R^{\uncontrol}/\gamma$, how the chemotactic performance and the optimal control function depend on the ratio between the time scale of the reference tumbling rate, $(R^{\uncontrol})^{-1}$, and the time scale of the utility discounting, $\gamma^{-1}$.
The timescale of utility discounting indicates how far into the future the optimization will take into account. 
From Fig. \ref{fig:optimal_tumble_rate}(C), we can see that the fold change in the optimal control function decreases with the increase of $\ndR$, indicating that the tumbling rate needs to be controlled gently to maximize the long-term utility.
Fig. \ref{fig:utility} shows that there is an optimal value of $\ndR$ that maximizes the net displacement.
The existence of the optimum may be understood as the compromise between speed and accuracy of decision-making under sensory noise.
If $\ndR$ is small, the estimation of swimming direction can be accurate because the noisy signal can be integrated for a long time before the next tumbling occurs. However if $\ndR$ is too small, the initiation of tumble may be deferred even when the swimming direction is not appropriate.

In summary, the chemotactic performance in terms of the net displacement along the ligand gradient increases at the expense of a large control cost when the SN ratio is high and when the optimal control function has a steep slope.
The net displacement also depends on the balance between the time scale for tumbling and that for performance evaluation.
The optimal control function is an increasing function of $\post$ in a wide range of parameters, and the steepness of its slope is modulated depending on the parameters $\ndb$ and $\ndR$ but almost independently of $\ndC$. 

\section{Summary and Discussion}
In this work, we derived the optimal sensing and motor-control strategy under sensory noise which characterizes the performance limit of general run-and-tumble chemotaxis.
We also investigated the connection between the derived optimal strategy and the standard biochemical model of \ecoli's signaling pathway. By explicitly considering motor control and its optimality, we verified that \ecoli's model and data can be related to the structure of the optimal strategy more naturally and reasonably.
This result reinforces the idea that \ecoli exploits the structure of the optimal strategy for attaining the performance limit\cite{nakamura2021connection}.

To verify the connection more quantitatively with other experimental data, we need to model the motor control in a more detailed way because the proposed optimality model cannot be compared directly with the experimental data of \ecoli's motor control.
Motor state of \ecoli is experimentally measured by the clockwise (CW) bias \cite{cluzel2000ultrasensitive}.
To define the CW bias, the duration of tumble should be finite, but we derived the optimality model under the approximation that the duration of tumble is infinitely small.
One way to alleviate this problem is to model the motor control as a stochastic switching between run and tumble \cite{vladimirov2008dependence,jiang2010quantitative,dufour2014limits}.
Then, we may extend the proposed model by considering a Markov chain with 4 states, $(\dir_{t},\tilde{\motor}_{t})$, where $\dir_{t}\in\{+1,-1\}$ is the swimming direction and $\tilde{\motor}_{t}\in\{\mathrm{run},\mathrm{tumble}\}$ is a motor state.
Such extension may enable us to verify further whether the optimality model can explain the experimental data of the motor control.

With the above extension of motor control, we may discuss the optimality of broader classes of exploratory behaviors observed in other organisms \cite{Stocker2635,gomez2012active}.
For example, \textit{Drosophila} larva shows a head casting behavior by which gradients can be sensed without moving \cite{gomez2012active}.
We may treat such behavior by adding a head cast to the motor state, $\tilde{\motor}$, and by modeling the velocity and sensing signal appropriately during the head cast.
It would be interesting to see how the optimal frequency and duration of the head cast depend on conditions such as the signal-to-noise ratio in run and head cast states.

Another possible direction is to extend sensory models to multiple ligands.
In this paper, we assumed that there is a single type of an attractant ligand, but there may be many types of ligands which work as attractants or repellents with different importance \cite{kalinin2009logarithmic,yang2015relation,zhang2019escape}.
We may incorporate sensing of multiple ligands by modeling the gradient of each ligand and extending the sensing signal, $\obs_{t}$, to a vector-valued Wiener process.
The importance of different ligands may temporally change depending on a cell's internal state such as nutrition requirements \cite{rengarajan2016olfactory}.
It would be interesting to consider not only the regulation of motor but also the integration of sensory information.
By changing the sensory integration based on nutrition requirements, a cell may focus attention on gradients of important nutrients.
Also, there are generally cross-talks between receptors.
By modeling the sensing signal appropriately, we may discuss the effect of  cross-talks on the performance of sensing and motor control.

In the extensions described so far, the estimated state was limited to finite states.
However, if we consider further extensions such as motion in a two-dimensional space, cells may need to estimate the direction as a continuous quantity to control its behavior optimally.
The posterior distribution of a continuous quantity cannot be represented by a finite vector in general, and it is difficult to implement its storage and update strictly in living organisms.
We may also need to solve the HJB equation of the value function which takes the posterior distribution as an argument, which is intractable.
To treat such an extension, we may need to consider other formulations such as the active inference framework which uses variational approximation of posterior distributions \cite{alexander2020learning}.

The combination of the optimal filtering and KL control for partially observed systems can work as a theoretical basis for all these extensions and also for analyzing the efficiency and optimality of various chemotactic phenomena.

\section*{Acknowledgements}
The first author is supported by JSPS Research Fellowship Grant Number 20J21362. This research is supported by JSPS KAKENHI Grant Numbers 19H05799 and by JST CREST JPMJCR2011.

\newpage


\bibliographystyle{apsrev4-2}
\providecommand{\noopsort}[1]{}\providecommand{\singleletter}[1]{#1}%

\appendix

\end{document}